\documentstyle[epsfig]{basi}
\begin{document}
\title[An efficient modulation scheme for dual beam polarimetry]{An Efficient Modulation Scheme for Dual Beam Polarimetry} 
\author[K. Nagaraju, K. B. Ramesh, K. Sankarasubramanian and K. E. Rangarajan]%
       {K. Nagaraju\thanks{e-mail:nagaraj@iiap.res.in}, K. B. Ramesh, 
        K. Sankarasubramanian$^\dagger$ and K. E. Rangarajan \\ 
        Indian Institute Astrophysics, Bangalore 560 034, India \\
        $\dagger$ISRO Satellite center, Bangalore 560094, India }
\maketitle
\label{firstpage}
\begin{abstract}
An eight stage balanced modulation scheme for dual beam
polarimetry is presented in this paper. The four Stokes parameters
are weighted equally in all the eight stages of modulation resulting
in total polarimetric efficiency of unity. The gain table error inherent
in dual beam system is reduced by using the well known beam swapping
technique. The wavelength dependent polarimetric efficiencies of
Stokes parameters due to the chromatic nature of the waveplates are
presented. The proposed modulation scheme produces better Stokes $Q$
and $V$ efficiencies for wavelengths larger than the design
wavelength whereas Stokes $U$ has better efficiency in the shorter
wavelength region. Calibration of the polarimeter installed as a
backend instrument of the Kodaikanal Tower Telescope is
presented. It is found through computer simulation that a $14\%$ sky
transparency variation during calibration of the polarimeter can
introduce $\approx 1.8\%$ uncertainty in the determination of its
response matrix.

\end{abstract}

\section{Introduction}
\label{sec:intro}
Polarimetric accuracy is one of the most important goals in modern
astronomy. It is limited since most optical elements encountered by
the light on its path from the source to the detector, can alter
its state of polarization (for eg. telescope optics, imaging system,
grating, etc). Apart from these, variation in sky transparency,
image motion and blurring due to the atmosphere  are a major concern
in high precision ground based solar polarimetry. The effect of
atmosphere, which is commonly known as seeing induced effect, can be
reduced by fast modulation schemes(Stenflo and Povel, 1985). The modulation
frequencies in these schemes are generally higher than seeing
fluctuations, which is $\approx 1kHz$(Stenflo and Povel, 1985 and Lites, 1987).
Large format CCDs, which are required to cover reasonable spectral
and spatial range, will pose difficulty in reading out the data at
$kHz$ speed. Stenflo and Povel(1985) proposed a scheme
whereby rapidly modulated signal is demodulated by optical means,
thereby avoiding the need to read the detectors at a rapid rate.
Lites(1987) has proposed a rotating waveplate modulation
scheme as an alternative to minimize the seeing induced cross-talk
among Stokes parameters. There he has shown that the faster the
rotation rate of the modulator, the lower the cross-talk among Stokes
parameters. And the seeing induced cross-talk levels of a dual beam
system are factors 3-5 smaller than those of a single beam system.
However, in dual beam system, the error introduced due to flat field
residual is a matter of concern in high precision polarimetry. A
possible solution to the above mentioned problems can be found by
using a mixed scheme in which spatial and temporal modulations are
performed(Elmore et al. 1992, Martinez Pillett et al. 1999 and
Sankarasubramanian et al. 2003). 
The gain table uncertainties are avoided using the beam swapping
technique(Donati et al. 1990, Semel et al. 1993 and Bianda et al. 1998)

A low cost dual beam polarimeter has been installed as a backend instrument for
the Kodaikanal Tower Telescope (KTT). Different modulation schemes
were studied and an optimum scheme is identified. The proposed scheme
requires eight stages of modulation of input light in order to obtain
the maximum polarimetric efficiency. Laboratory experiments have been
performed to verify the theoretical understanding of the proposed
scheme. The studies are extended to other wavelengths apart from the
design wavelength of $\lambda6300$.

The outline of this paper is as follows. The proposed eight stage
modulation scheme for the measurement of general state of
polarization is discussed in section(2). Wavelength dependence of the
efficiency of the polarimeter in measuring Stokes parameters is
presented in section(3). In section(4), the performance of the
polarimeter at KTT is presented.

\section{Proposed Modulation Scheme}
\label{sec:1}

A zero-order quarter wave (R1) and a zero-order half wave (R2)
retarders at $\lambda6300$ are chosen as the modulators for the proposed
polarimeter. R1 and R2 are the first and second elements of the
polarimeter, as seen by the incoming light, followed by a polarizing
beam splitter cube (PBS). The PBS has an extinction ratio $> 10^3$
with respect to a polarizing Glan-Thomson prism (GTP). A simplest way
of measuring Stokes parameters is to use a half waveplate (HWP) along
with PBS for linear polarization measurement and a quarter waveplate
(QWP) along with PBS for circular polarization measurement
(Bianda et al. 1998). However, this way of modulation will introduce
a possible differential optical aberrations between the linear and
circular polarization measurements due to different optical elements
encountered by the light. Using both the waveplates during all stages
of measurements or using a single retarder with an appropriate
retardance can avoid the differential aberrations(Lites, 1987 and Elmore et al. 1992). The modulation scheme presented here uses both R1
and R2 in all stages of measurements.

\subsection{Modulation}
The input polarization is modulated on to intensity by using the 
waveplate orientations given in Table.1.
 
\begin{table}[h]
\label{tab:orient}
\centering
\begin{center}
\begin{tabular}{lll}
\hline
Modulation & Orientation of& Orientation of \\
stage & QWP(R1) & HWP(R2) \\\hline
1 & 22.5 & 0 \\
2 & 22.5 & 45 \\
3 & 67.5 & 45 \\
4 & 67.5 & 90 \\
5 & 112.5 & 90 \\
6 & 112.5 & 135 \\
7 & 157.5 & 135 \\
8 & 157.5 & 180 \\ \hline
\end{tabular}
\end{center}
\caption{ Orientation of Waveplates for different stages of modulation
expressed in degrees.}
\end{table}

The modulated intensities $\vec{I}^{\pm}=(I_1^{\pm}, I_2^{\pm}, I_3^{\pm}, I_4^{\pm}, I_5^{\pm}, I_6^{\pm}, I_7^{\pm}, I_8^{\pm})^T$, where $T$
represents transpose operator, can be written in terms of input
Stokes parameters as
\begin{equation}
\label{eq:mod1}
\vec{I}^{\pm} = g^{\pm} {\bf O}^{\pm} \vec{S}_{in}.
\end{equation}
Where $\pm$ indicate the two orthogonally polarized beams emerging
out of the polarimeter respectively. ${\vec S}_{in} = [I, Q, U, V]^T$ is
the input Stokes vector to the polarimeter. Here, the standard
definition of the Stokes vector is used with $I$ representing the
total intensity, $Q$ and $U$ representing the linear polarization
state and $V$ representing the  circular polarization state. The
multiplication factor of the two orthogonally polarized beams
$g^{\pm}$, known as the gain factor, is a product of transparency of
the corresponding optical path and the detector gain factor. 
The analyser Mueller matrices of respective beams can be obtained
by multiplying the Mueller matrices of 
retarders(${\bf M}_{R1}$ and ${\bf M}_{R2}$) and PBS(${\bf M}^{\pm}_{P}$)
in the order ${\bf M}^{\pm}_{P}{\bf M}_{R2}{\bf M}_{R1}$ 
(del Toro Iniesta, 2003 and Stenflo, 1994).
The modulation matrices ${\bf O}^{\pm}$ are constructed by arranging the first
row of the analyser matrix of the respective beam 
for each of the measurement steps(see del Toro Iniesta, 2003 for details).

The theoretical modulation matrices ${\bf O}^{\pm}$ at the design wavelength are 
given below.

\[ {\bf O}^{\pm} = 0.5\left( \begin{array}{cccc}
1.0 & \pm 0.5 &  \mp 0.5  & \pm 0.707\\
1.0 & \mp 0.5 &  \pm 0.5  & \mp 0.707\\
1.0 & \mp 0.5 &  \mp 0.5  & \mp 0.707\\
1.0 & \pm 0.5 &  \pm 0.5  & \pm 0.707\\
1.0 & \pm 0.5 &  \mp 0.5  & \mp 0.707\\
1.0 & \mp 0.5 &  \pm 0.5  & \pm 0.707\\ 
1.0 & \mp 0.5 &  \mp 0.5  & \pm 0.707\\
1.0 & \pm 0.5 &  \pm 0.5  & \mp 0.707
\end{array} \right). \]

It is to be noted here that the four Stokes parameters are modulated
on to intensity in all the eight stages of measurements. Also that in
all the eight stages of measurements, each Stokes parameter is
weighted equally. Matrices ${\bf O}^{\pm}$ show that the alternate measurements
are obtained by swapping the orthogonally polarized beams (seen as
sign change in the corresponding Stokes parameters). 

The maximum efficiencies of the modulation scheme
(see del Toro Iniesta and Collados, 2000 and del Toro Iniesta, 2003 for
details) in measuring Stokes $I$, $Q$, $U$, $V$ are 
1.0,~0.5,~0.5,~0.707 respectively, at the design wavelength. 
The total polarimetric efficiency is $\sqrt{0.5^2+0.5^2+0.707^2} = 0.9999$, 
which is close to unity as expected since the absolute values of all the
elements in a given column are same(del Toro Iniesta, 2003). 
The values of matrix elements ${\bf O}^{\pm}$ are  not the same 
at different wavelengths and hence the maximum efficiencies of
modulation scheme in measuring Stokes QUV are different.
However, the total polarimetric efficiency will remain close to
unity. 

For comparison, the maximum efficiencies of some of the well known
polarimeters are given below:
ASP-(1.0,~0.546,~0.41,~0.659),
ZIMPOL-(1.0,~0.474,~0.467,~0.534), 
TIP-(1.0,~0.617,~0.41,~0.659),
POLIS-(1.0,~0.494,~0.464,~0.496).
In the above examples, only TIP has a total polarimetric efficiency
close to unity.
\subsection{Demodulation}

The demodulation of the input Stokes parameters from the modulated
intensities involve the following steps.
As a first step, the signal vectors  $\vec{S}^{\pm}$ 
(Gandorfer, 1999 and Stenflo, 1984) of the orthogonally polarised 
beams are constructed from the modulated intensities (Eq. \ref{eq:mod1}) 
using the equation,
\begin{equation}
\label{eq:dmod1}
\vec{S}^{\pm} = {\bf D} \vec{I}^{\pm}/8.
\end{equation}
Where the matrix ${\bf D}$ is defined as,
\[ {\bf D} = \left( \begin{array}{cccccccc}
~~~1.0 & ~~~1.0 & ~~1.0 & 1.0 & ~~1.0 & ~~1.0 & ~~1.0 & ~1.0 \\
 ~~~1.0 &  -1.0 & -1.0 &  1.0 &  ~~1.0 &  -1.0 & -1.0 &  1.0  \\
-1.0 &  ~~1.0 &  -1.0 &  1.0 &  -1.0 & ~~1.0 & -1.0 &  1.0 \\
 ~~1.0 &  -1.0 & -1.0 & 1.0 & -1.0 & ~~1.0 & 1.0 & -1.0 \\
\end{array} \right). \]
One can see from the matrix ${\bf D}$ that, to derive signal vectors
all the eight stage intensity measurements are considered.
Hence the final derived input Stokes parameters will be well balanced with
respect to changes, if there are any,  during measurements in the input Stokes
parameters. Also, the final derived Stokes parameters will be an average
over the time taken for the eight stages of modulation.

The second step involves combining the signal vectors of the 
orthogonally polarized beams after correcting for the gain factors $g^{\pm}$.
Gain table corrections can be done either by regular flat field procedure 
or normalizing the elements of signal vectors to their respective first element 
(i.e. $\vec{S}^{\pm}/S^{\pm}(0)$). 
Since the regular flat field procedure limits the polarimetric precision and 
to make use of the advantage of the beam swapping technique incorporated in 
the modulation scheme, second method is used to derive the combined signal
vector. The signal vector($\vec{S}'$) of the combined beam can be written as

\begin{eqnarray} 
\label{eq:dmod2}
S'(0) &=& S^+(0)+S^-(0) \\\nonumber
S'(1) &=& S^+(1)/S^+(0) - S^-(1)/S^-(0) \\\nonumber
S'(2) &=& S^+(2)/S^+(0) - S^-(2)/S^-(0) \\\nonumber
S'(3) &=& S^+(3)/S^+(0) - S^-(3)/S^-(0) \\\nonumber
\end{eqnarray}
where $S'(i)$, $i=0,1,2,3$, are the elements of $\vec{S}'$.
Similarly $S^{\pm}(i)$ are defined. The indices $i=0,1,2,3$ are correspond
to the Stokes parameters I, Q, U, V respectively.

However, the flat field corrections are essential for the total intensity.
With the definitions of Eq.(\ref{eq:dmod2}),  the signal vector of the 
combined beam can be written in terms of the input Stokes vector as
(Gandorfer, 1999  and Stenflo, 1984),
\begin{equation}
\label{eq:dmod3}
\vec{S}'={\bf M}\vec{S}'_{in}.
\end{equation}
Where $\vec{S}'_{in}=[I, Q/I, U/I, V/I]^T$ is the input Stokes vector 
and {\bf M} is a $4\times4$ matrix known as the response matrix of the 
polarimeter. 
The theoretical response matrix at the design wavelength is given by
\[ {\bf M}=\left( \begin{array}{cccc}
1       & 0  &  0 & 0\\
0 & 0.5 &  0 & 0  \\
0 & 0   &  0.5  & 0 \\
0 & 0   &  0&  0.707
\end{array} \right). \]

The third and final step involves obtaining the input 
Stokes vector($\vec{S}'_{in}$) from  Eq.(\ref{eq:dmod3}). 
During the observations, the response matrix is obtained
using a polarimetric calibration procedure which will be detailed in
Section 4.

\subsection{Polarimetric Efficiency}
The efficiency of the polarimeter in measuring respective
Stokes parameter is defined as (Beck et al. 2005)
\begin{equation}
\label{eq:effdef1}
\epsilon_i = \sqrt{\sum_{j=1,4}{\bf M}_{ji}^2}.
\end{equation}
where i = 0, 1, 2, 3 corresponds to I, Q, U, V respectively.
Since the response matrix ${\bf M}$ in Eq.(\ref{eq:dmod3}) is diagonal, 
the efficiency in the Eq. (\ref{eq:effdef1}) will be simplified 
(using Eq. \ref{eq:dmod3}) to 
\begin{equation}
\label{eq:effdef2}
\epsilon_i = |S'(i)/S'_{in}(i)|.
\end{equation}

We would like to note here that the response matrix ${\bf M}$ in 
Eq.(\ref{eq:dmod3}) is diagonal at all the wavelengths considered here.
However, at the wavelengths away from the design wavelength, 
the signals $S^{\pm}(0)$ in Eq.(\ref{eq:dmod1}) are not just proportional to 
input Stokes I but with a small contribution from the input Stokes Q. 
If the signal vectors($\vec{S}^{\pm}$) of orthogonally polarised beams
are combined without normalising to their respective Stokes I signal
then this cross-talk term will not appear in the signal vector($\vec{S}'$) of 
the combined beam. But, the flat fielding is essential to remove the
gain factors $g^{\pm}$.
In this paper the signal vectors are combined in such a way that the
Stokes QUV signals are normalised to Stokes I signal in order to remove
the gain factors. 
This results in an over estimation of efficiency $\epsilon_Q$.
This over estimation is about 1.8\% at $\lambda4500$, 0.15\% at $\lambda5000$,
0.65\% at $\lambda5500$, 0.24\% at $\lambda5890$, 0.1\% at $\lambda6563$
and 0.6\% at $\lambda7000$. These wavelengths are chosen because, the
efficiencies are measured at these wavelengths through the experiment
presented in the next section.
In the regular solar observations the Stokes Q signal is smaller 
than Stokes I at least by an order of magnitude,
the cross-talk from Q to I will  be often negligible.

\section{Laboratory Experiment}  
\label{sec:labexp}
The efficiency of the polarimeter is different for different
wavelengths due to the chromatic nature of the retarders used in the
polarimeter. If the polarimeter is used at different wavelengths then
it is important to understand its performance at the desired
wavelength. 
The variation of the efficiency factor for different input Stokes
parameters is studied by carrying out a few laboratory experiments.

\begin{figure}
\centering
\includegraphics[width=14cm,height=4cm]{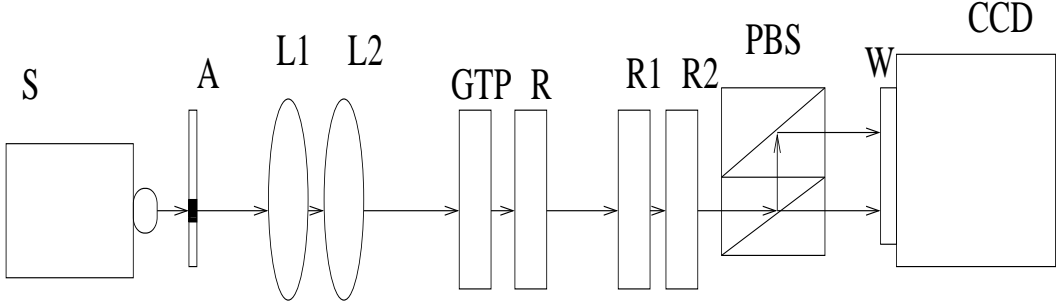}
\caption{Block diagram of an experimental setup used to study the
polarimetric efficiency.
The symbols in the figure are S-monochromator, A-mount holding a
$1mm$ square aperture, L1 and L2-lenses(f=25cm), GTP-Glan Thomson prism used to
produce linear polarization, R-QWP used along with GTP to
produce circular polarization, R1 and R2- Half and Quarter waveplates
which forms a part of the polarimeter, PBS-polarizing beam splitter,
W- CCD window.}
\label{fig:expsetup}
\end{figure}

The experimental setup is shown in Fig.\ref{fig:expsetup}. The light
from the monochromator was set at the desired wavelength and then
passed through a $1mm$ rectangular aperture. This rectangular
aperture was imaged on to a CCD detector using a two lens system with
an effective focal length of $12.5cm$. The polarimeter optics were
placed between the lens and the detector. The first in the light
beam is the QWP (R1), followed by the HWP (R2) and the PBS. The
retarders (R1 and R2) of the polarimeter were mounted on two
different rotating stages whose rotational accuracy is $0.1^o$. A
known state of polarization was produced using the GTP and
calibration retarder(CU) R. Stokes $Q$ and $U$ were produced by using only
the GTP where as both GTP and R were used to produce Stokes $V$.
(The CU retarder is a zero order chromatic waveplate which acts as
a quarter waveplate at $\lambda6300$. This is the same quarter waveplate
which was earlier used in the single beam polarimeter installed at Kodaikanal
by Sankarasubramanian(2000) and the characteristics of the waveplate
is presented there. The retardance of the CU retarder at other wavelengths
is calculated by the well known  linear relation, see for eg. in Hetch, 2002).
Eight measurements, for each input Stokes parameter, were performed
by positioning the retarders (R1 and R2) at different angles as shown
in Table.\ref{tab:orient}.

The measured data were first corrected for dark current. 
Then the Stokes signal vectors correspond to the orthogonally polarised
beams were obtained from the measured intensities and the
matrix {\bf D} using  Eq.(\ref{eq:dmod1}). 
The corresponding Stokes QUV signals were normalised to the 
respective Stokes I signal as in Eq.(\ref{eq:dmod2}). 

We noted in the section 2.3 that the theoretical response matrix
of the polarimeter presented in this paper is diagonal at all
the wavelengths considered.  However, in practice there will
be off-diagonal elements which are nothing but the cross-talk
among Stokes parameters. But, in the experiment performed
to study the polarimetric efficiency, the measured cross-talk
terms are small. Since the cross-talk terms are small,
the simplified definition of Eq.(\ref{eq:effdef2}) is
used to calculate the efficiency.

\begin{figure}
\centering
\includegraphics[width=12cm,height=16cm]{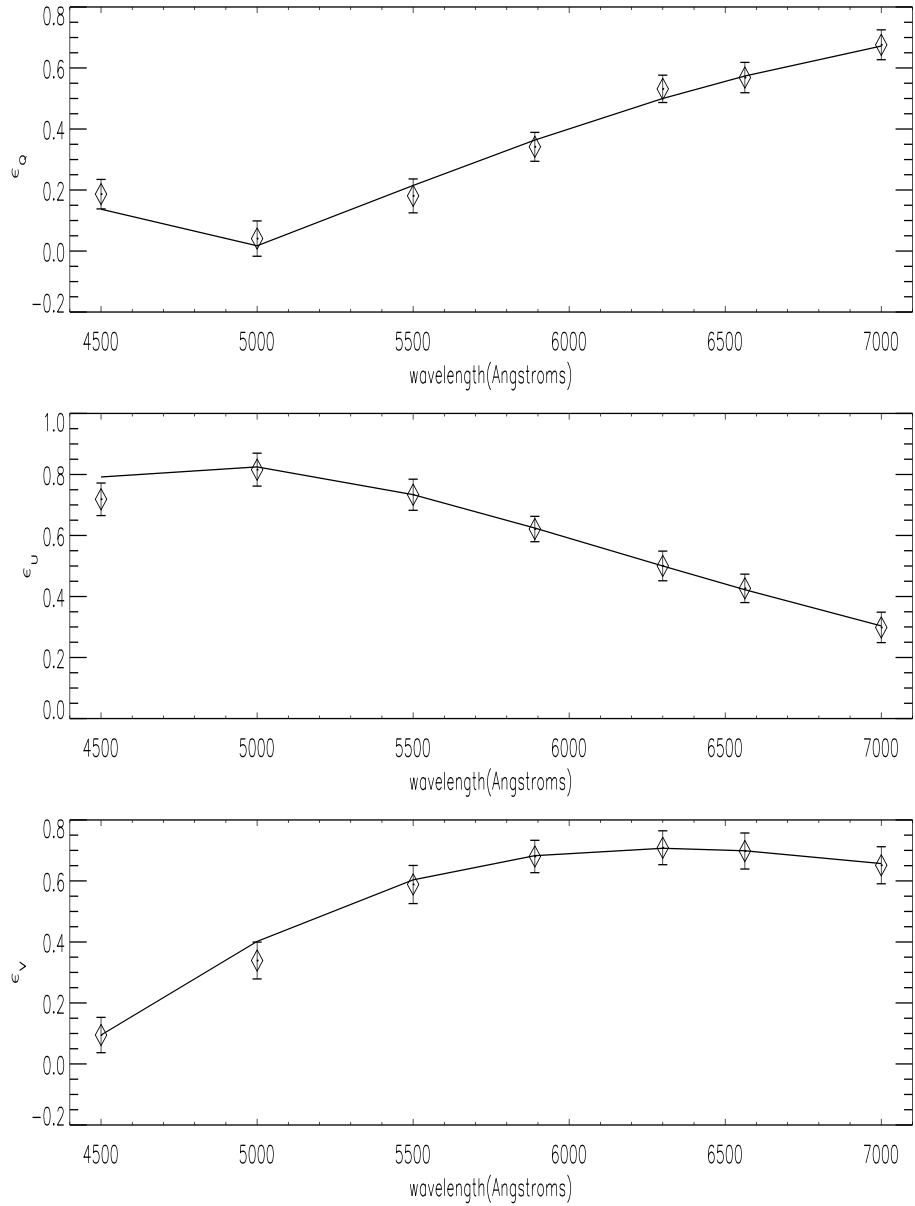}
\caption{Plots of polarimetric efficiency of Stokes $Q$, $U$ and $V$
parameters as a function of wavelength. The solid curves are
theoretical values where as the diamond symbols correspond to the measured
values.
The error bars shown in these plots are ten times the obtained noise rms
values.}
\label{fig:efficiency}
\end{figure}

Plots of polarimetric efficiency in measuring Stokes $Q$, $U$ and
$V$ as a function of wavelength are shown in
Fig.\ref{fig:efficiency}. The diamond symbols shown in the plots are the
experimental values and the solid lines are theoretical curves. It is
clear from the plot that the experimental values closely resemble the
theoretical predictions. From this figure it can be concluded that
$Q$ and $V$ are measured with better efficiencies in longer
wavelength region compared to the design wavelength, where as $U$ is
measured better at shorter wavelengths.

\section{Calibration of the Polarimeter Installed at KTT}
   
The calibration of the polarimeter, which uses the modulation
scheme  presented in section 2, installed as a backend instrument at 
KTT is discussed in this section.
KTT is a three mirror coelostat system for solar observations(Bappu, 1967).
It is equipped with a Littrow mount spectrograph
using a grating of 600 lines per mm. The polarimeter setup is placed
in the converging beam (f/90) of the telescope before the
spectrograph slit. The retarders are mounted on rotating stages which
can be rotated from $0$ to $360^o$ in steps of $22.5^o$. PBS is fixed
in position with one of its optic axes parallel to the slit
direction. A linear polarizer, called as a compensator, is placed
behind the PBS to compensate for the differential grating efficiency
of the two orthogonally polarized beams. The compensator is oriented
at $45^o$ to the grating grooves to make the efficiency of the
orthogonal beams nearly equal. Ideally a QWP or a HWP will be
preferred as the compensator. But, linear polarizer is used to make
use of the polarimeter at other wavelengths of interest.

To calibrate the polarimeter, a calibration unit consisting of a
linear polarizer followed by a quarter waveplate at $\lambda 6300$,
was used. The optical axis of the linear polarizer was aligned with
one of the optic axes of the PBS. Keeping the polarizer at this
orientation, the linear retarder was rotated from $0^o$ to $180^o$
in steps of $15^o$ and hence producing 13 input states of polarisation.
Out of 13 states of polarisation, only 11 are different because the
polarisation states correspond to the retarder orientation
$0^o$, $90^o$ and $180^o$ are essentially the same.
The corresponding input Stokes parameters and measured Stokes signals
(Eq. \ref{eq:dmod2}) are arranged in a $13\times 4$ matrix 
form and solved for response matrix of the polarimeter as 
follows(Beck et al. 2005). If
${\bf S}_{in}^c$ represents the $13\times 4$ input Stokes matrix 
and ${\bf S}_{op}^c$ represents the measured 
$13\times 4$ signal matrix then,

\begin{equation}
\label{eq:calibrespmat1}
{\bf S}_{op}^c = {\bf S}_{in}^c {\bf M}^T.
\end{equation}

The response matrix of the polarimeter setup (${\bf M}$) is solved by
defining $({\bf S}_{in}^c)^T{\bf S}_{op}^c = ({\bf S}_{in}^c)^T{\bf S}_{in}^c{\bf M}^T = {\bf S}{\bf M}^T$ as (see Beck et al. 2005 for details)

\begin{equation}
\label{eq:matmod}
{\bf M}^T = {\bf S}^{-1} ({\bf S}_{in}^c)^T{\bf S}_{op}^c.
\end{equation}
Where, ${\bf S} = ({\bf S}_{in}^c)^T {\bf S}_{in}^c$. The structures of
${\bf S}_{in}^c$ and ${\bf S}_{op}^c$ are given in the appendix for the
sake of clarity.

\subsection{ Measurement of response matrix of the polarimeter}

The calibration data were obtained using the procedure as explained
in the beginning of this section. Data was subjected to the
standard dark and flat corrections(for eg. see Beck et al. 2005). 
The response matrix of the
polarimeter setup has been derived using the $13 \times 4$ input Stokes
matrix(${\bf S}^c_{in}$) constructed out of input Stokes parameters and
$13 \times 4$ signal matrix(${\bf S}^c_{op}$) constructed out of 
measured Stokes signals using Eq.(\ref{eq:matmod}). A typical
derived response matrix(${\bf M}$) of the polarimeter setup is,

\[ \left( \begin{array}{cccc}
1       & -0.0066  &  ~0.0384 & ~~0.0461\\
~~0.0048 & ~~0.5643 &  ~0.0565 & -0.0016  \\
~~0.0074 & -0.0084  &  ~0.4173 & -0.0065 \\
-0.0053 & -0.0037  &  ~0.0248 &  ~~0.6835
\end{array} \right). \]

The wavelength of observation is in the continuum of $\lambda6563$
wavelength region. The fit error of the second, third and fourth
column of the response matrix, which are nothing but the errors in the 
determination of Stokes $QUV$, are 0.0044, 0.0048 and 0.0022 respectively. The
corresponding noise rms of the measurements are 0.0017, 0.0018 and
0.0023 respectively.

The observed efficiency(Eq.\ref{eq:effdef1}) of the polarimeter in measuring 
Stokes $QUV$ are 0.5644, 0.4228 and 0.685086 respectively, which are close to 
theoretically expected values of  $0.573$, $0.423$ and $0.698$ at 
this wavelength.
In an ideal case, off-diagonal elements of the response matrix ${\bf M}$
are expected to be zero. However, in practice telescope induced
cross-talk and the variation in the sky transparency can influence
the calibration of the polarimeter. Sky transparency variation means
that the variation in the input intensity to the telescope. To take
into account the telescope induced cross-talk, telescope model of KTT
originally developed by Balasubramaniam et al.(1985) and
later modified by Sankarasubramanian(2000) is used. A
computer simulation has been performed to understand the effect of sky
transparency variation on the calibration of the polarimeter. From
this simulation it is found that the maximum cross-talk produced
among Stokes parameters is $\approx 1.8\%$ for a sky transparency
variation of $14\%$ within the eight stages of measurement. In the
actual measurements for this calibration, the intensity variation
is of the order of $14\%$. The origin of most of the off-diagonal
elements ${\bf M}$, can be explained based on the sky transparency
variation. However, there are off-diagonal elements which are 
larger than the $1.8\%$ expected due to the sky transparency
variation. An off set angle of $\approx -1.5^o$ in the HWP is
required to produce the observed cross-talk from $U$ to $Q$. However,
the origin of cross-talk from $U$ to $V$ has not been traced out.

\section{Conclusions}

An eight stage modulation scheme to measure the general state of
polarization is presented here. Beam swapping technique is
incorporated in this scheme, which helps in alleviating the gain
correction errors. The total polarimetric efficiency is close to
unity as the Stokes parameters are weighted equally in all the stages
of modulation. The final Stokes parameters are demodulated using all
the stages of intensity measurements. Hence, the derived input Stokes
parameters are equally weighted time averaged quantities over the
time of measurement.

Since the retarders used in the polarimeter are chromatic, the
efficiency of the polarimeter in measuring Stokes $QUV$ is wavelength
dependent. The laboratory experiments performed to study the
wavelength dependence of efficiency of the polarimeter confirms the
theoretical expectations.

It is found through computer simulation that a $14\%$ sky
transparency variation can cause $\approx 1.8\%$ uncertainty in the
elements of the polarimetric response matrix during its
calibration, for the modulation/demodulation scheme presented here.
The non-zero values of the off-diagonal elements are not a serious 
concern if those values do not change drastically in short time
scales. During any solar polarimetric measurements, data for the
calibration of the polarimeter are taken at least once a day. 
Calibration of the polarimeter are carried out on a few days over a
period of 10-day and the response matrix derived over this period
did not show any appreciable variations. The variations in the
off-diagonal elements are less than the fit errors ($<$ 0.5\%).
The polarization signals observed on the Sun is always less than 40\%
and hence an uncertainty of 0.5\% in the calibration will produce
an inaccuracy of 0.2\% in the polarization signals.

	The measured total polarimetric efficiency of the polarimeter 
installed at KTT is $\approx 0.986$ at $\lambda6563$ wavelength region 
which is better than some
of the polarimeters such as ZIMPOL(0.72), ASP(0.88), TIP(0.92) and POLIS(0.84).

\section*{Acknowledgments}
We would like to thank the anonymous referee for the useful suggestions which 
made the contents of the paper clearer.
We thank B. R. Prasad and Ravinder Kumar Baynal for providing some of the
laboratory equipments necessary for the experiment and P. K. Mahesh for
his help in procuring the laboratory equipments. We thank P. U. Kamath for
his help in mechanical design and fabrication of the polarimeter. The help of
P. Devendran and P. Hariharan during observations are thankfully acknowledged.

\begin{center}
{\bf Appendix} 
\end{center}
The structure of the measured Stokes signal matrix in 
Eq. (\ref{eq:calibrespmat1}) is given by
\[ {\bf S}_{op}^c=\left( \begin{array}{cccc}
S_{op}^1(0)       & S_{op}^1(1)  &S_{op}^1(2)  &S_{op}^1(3) \\
S_{op}^2(0)       & S_{op}^2(1)  &S_{op}^2(2)  &S_{op}^2(3) \\
S_{op}^3(0)       & S_{op}^3(1)  &S_{op}^3(2)  &S_{op}^3(3) \\
S_{op}^4(0)       & S_{op}^4(1)  &S_{op}^4(2)  &S_{op}^4(3) \\
S_{op}^5(0)       & S_{op}^5(1)  &S_{op}^5(2)  &S_{op}^5(3) \\
S_{op}^6(0)       & S_{op}^6(1)  &S_{op}^6(2)  &S_{op}^6(3) \\
S_{op}^7(0)       & S_{op}^7(1)  &S_{op}^7(2)  &S_{op}^7(3) \\
S_{op}^8(0)       & S_{op}^8(1)  &S_{op}^8(2)  &S_{op}^8(3) \\
S_{op}^9(0)       & S_{op}^9(1)  &S_{op}^9(2)  &S_{op}^9(3) \\
S_{op}^{10}(0)       & S_{op}^{10}(1)  &S_{op}^{10}(2)  &S_{op}^{10}(3) \\
S_{op}^{11}(0)       & S_{op}^{11}(1)  &S_{op}^{11}(2)  &S_{op}^{11}(3) \\
S_{op}^{12}(0)       & S_{op}^{12}(1)  &S_{op}^{12}(2)  &S_{op}^{12}(3) \\
S_{op}^{13}(0)       & S_{op}^{13}(1)  &S_{op}^{13}(2)  &S_{op}^{13}(3) \\
\end{array} \right). \]

And that of input Stokes matrix is given by
\[ {\bf S}_{in}^c=\left( \begin{array}{cccc}
S_{in}^1(0)       & S_{in}^1(1)  &S_{in}^1(2)  &S_{in}^1(3) \\
S_{in}^2(0)       & S_{in}^2(1)  &S_{in}^2(2)  &S_{in}^2(3) \\
S_{in}^3(0)       & S_{in}^3(1)  &S_{in}^3(2)  &S_{in}^3(3) \\
S_{in}^4(0)       & S_{in}^4(1)  &S_{in}^4(2)  &S_{in}^4(3) \\
S_{in}^5(0)       & S_{in}^5(1)  &S_{in}^5(2)  &S_{in}^5(3) \\
S_{in}^6(0)       & S_{in}^6(1)  &S_{in}^6(2)  &S_{in}^6(3) \\
S_{in}^7(0)       & S_{in}^7(1)  &S_{in}^7(2)  &S_{in}^7(3) \\
S_{in}^8(0)       & S_{in}^8(1)  &S_{in}^8(2)  &S_{in}^8(3) \\
S_{in}^9(0)       & S_{in}^9(1)  &S_{in}^9(2)  &S_{in}^9(3) \\
S_{in}^{10}(0)       & S_{in}^{10}(1)  &S_{in}^{10}(2)  &S_{in}^{10}(3) \\
S_{in}^{11}(0)       & S_{in}^{11}(1)  &S_{in}^{11}(2)  &S_{in}^{11}(3) \\
S_{in}^{12}(0)       & S_{in}^{12}(1)  &S_{in}^{12}(2)  &S_{in}^{12}(3) \\
S_{in}^{13}(0)       & S_{in}^{13}(1)  &S_{in}^{13}(2)  &S_{in}^{13}(3) \\
\end{array} \right). \]
Where $S_{op}^j(i)$ and $S_{in}^j(i)$, $j=1.....13$ 
correspond to 13 orientations of the calibration retarder and 
$i = 0,1,2,3$ corresponds to Stokes I,Q,U,V are the measured Stokes signals and
input Stokes parameters respectively.

\end{document}